# Machine learning assisted quantum super-resolution microscopy

Zhaxylyk A. Kudyshev[1,2], Demid Sychev[1,2], Zachariah Martin[1,2], Simeon I. Bogdanov[3,4], Xiaohui Xu[1,2], Alexander V. Kildishev[1], Alexandra Boltasseva[1,2], Vladimir M. Shalaev[1,2]

[1] *School of Electrical and Computer Engineering, Birck Nanotechnology Center and Purdue Quantum Science and Engineering Institute, Purdue University, West Lafayette, IN, 47907, USA*
[2] *The Quantum Science Center (QSC), a National Quantum Information Science Research Center of the U.S. Department of Energy (DOE), Oak Ridge, TN, 37931, USA*
[3] *Department of Electrical and Computer Engineering, University of Illinois at Urbana-Champaign, Urbana, IL, 61801, USA*
[4] *Nick Holonyak, Jr. Micro and Nanotechnology Laboratory, University of Illinois at Urbana-Champaign, Urbana, IL, 61801*, USA

*shalaev@purdue.edu*

**Abstract:** One of the main characteristics of optical imaging systems is the spatial resolution, which is restricted by the diffraction limit to approximately half the wavelength of the incident light. Along with the recently developed classical super-resolution techniques, which aim at breaking the diffraction limit in classical systems, there is a class of quantum super-resolution techniques which leverage the non-classical nature of the optical signals radiated by quantum emitters, the so called antibunching super-resolution microscopy. This approach can ensure a factor of $\sqrt{n}$ improvement in the spatial resolution by measuring the $n$-th order autocorrelation function. The main bottleneck of the antibunching super-resolution microscopy is the time-consuming acquisition of multi-photon event histograms. We present a machine learning-assisted approach for the realization of rapid antibunching super-resolution imaging and demonstrate 12 times speed-up compared to conventional, fitting-based autocorrelation measurements. The developed framework paves the way to the practical realization of scalable quantum super-resolution imaging devices that can be compatible with various types of quantum emitters.

Introduction

Due to the wave nature of light, the spatial resolution of conventional far-field microscopes is fundamentally limited by the diffraction limit to approximately half the wavelength of the incident light, known as the Rayleigh criteria [1] or Abbe limit [2]. Far-field super-resolution microscopy (SRM) techniques that aim at overcoming the diffraction limit could greatly impact the fields of biology, physics and chemistry, as well as device engineering, semiconductor industry, and could lead to novel applications [3–8]. The developed SRM techniques typically break one or more of the underlying fundamental assumptions on the nature of light-matter interaction within the optical system, under which the diffraction limit is derived. Specifically, it is assumed that the illumination intensity is homogenous, the optical response of the stationary object is linear, and all the optical fields in the system are classical. Recently, a plethora of novel super-resolution techniques, including stimulated emission depletion [9], structured illumination microscopy [10], photoactivated localization microscopy [11], and stochastic optical reconstruction microscopy [12] has been developed. All the aforementioned techniques are realized within classical optical systems via breaking the homogeneity, linearity, or stationarity assumptions.

Another promising route in the realization of SRM techniques is to take into account the quantum nature of light [13–16]. Recently, several quantum schemes, utilizing multimode squeezed light [17] and generalized quantum states [18] have been proposed. These approaches use complex quantum states of light as an illumination source, which demand highly efficient, deterministic sources of such quantum photons or entangled photon pairs. In contrast, several SRMs have been developed by relying on the quantum nature of the light emitted by the object itself. This approach, originally proposed by Hell et al [19], is based on fact that some quantum sources of light produce emission with sub-Poissonian temporal photon statistics, which can be analyzed by measuring the autocorrelation function of the emission. By analyzing the $n$-th order autocorrelation function at zero time delay $g^{(n)}(t=0)$ of nonclassical light emitted from a point source, it is possible to reduce the size of the effective point spread function by a factor of $\sqrt{n}$ [20–22].

The antibunching-based SRM can be coupled with a classical approach to further improve the resolution of the imaging system. By combining image scanning microscopy with the measurement of the second-order quantum photon correlation, a spatial resolution of four times beyond the diffraction limit was achieved [23] with only a modest hardware overhead compared to regular confocal scanning microscopy. This combination makes the antibunching-based SRM technique a very attractive platform for imaging quantum light sources, as these are typically analyzed using confocal scanning microscopy. The main bottleneck of this framework is the time required for the acquisition of the time-resolved photon statistics needed to accurately determine the values of the autocorrelation function at zero delay. This accuracy depends on the number of registered correlated photon detection events. The time requirement scales up exponentially with the increasing order of the autocorrelation function. Hence, in order to realize scalable and practical antibunching-based SRM, one needs to develop a fast and precise approach to determine g$^{(n)}$(0).

Recently, convolutional neural networks (CNNs) enabled the rapid classification of quantum emitters depending on whether g$^{(2)}$(0) is above or below a given threshold value based on sparse autocorrelation function measurements [24]. Leveraging on these results, we present a CNN-based regression model that allows an accurate estimation of the $g^{(2)}(0)$ value based on sparse data. Using the developed CNN model, we reduced the acquisition time in the antibunching-based scanning SRM technique by 12 times, thus marking an important step towards the practical realization of scalable quantum super-resolution imaging devices.

## 2. Machine learning assisted antibunching super-resolution microscopy

The antibunching SRM technique relies on the detection of quantum correlations in the signal radiated by quantum emitters, which allows for a gain in the spatial resolution of a factor of $\sqrt{n}$ by measuring $n$-th order autocorrelation function [21]. This fact can be understood by conducting the Gedanken experiment first proposed by Hell et al [19]. In the case of a hypothetical emitter that emits photons by pairs, an improvement in resolution can be theoretically obtained by sending each of the two photons to a separate camera. Since the two cameras will record two independent point-spread function (PSFs) estimates, the spatial resolution can be improved by a factor of $\sqrt{2}$ via simple multiplication. However, instead of requiring the emitter to emit pairs of photons, one can acquire the same amount of information by assessing an absence of the two-photon correlation in single photon emission by measuring the second-order autocorrelation function. Furthermore, one can achieve an arbitrarily high improvement in resolution by measuring higher-order correlations in the emission of a single photon emitter. In the most general form, the intensity distribution of the super-resolved image based on antibunching SRM $G^{(n)}(x,y)$ can be obtained via retrieving spatial distributions of the $n$-th order autocorrelation function at zero time delay $g^{(n)}(x,y,\tau=0)$ and the number of detected photons $\widetilde{N}(x,y)$ [21]:

$$G^{(n)}(x,y) \sim \langle \widetilde{N}(x,y) \rangle^n \sum_{i=1}^{i=i_{max}} c_i \chi_i, \qquad (1)$$

here $\langle \widetilde{N}(x,y) \rangle$ is the average number of detected photon from a given point $(x,y)$ of the sample; $\chi_i$ is a function of the product $g^{(j_1)}(x,y,0)g^{(j_2)}(x,y,0)\ldots g^{(j_l)}(x,y,0)$, where $i_{max}$ is the number of ordered combinations, fulfilling the condition $\sum_{k=1}^{l} j_k = n$. For example, for $n=2$ case, Eq.(2) takes the following simple form [21]:

$$G^{(2)}(x,y) \sim \langle \widetilde{N}(x,y) \rangle^2 \big(1 - g^{(2)}(x,y,0)\big). \qquad (2)$$

The most commonly used approach for retrieving the $g^{(2)}(0)$ value is a Hanbury-Brown-Twiss (HBT) interferometry measurement, composed of a beam-splitter directing the emitted light to two single-photon detectors connected to a correlation board (Figure 1a). The correlation board registers events consisting of pairs of detector clicks. It then arranges these events into a histogram as a function of the time delay $\tau$ between the clicks, which can be used for the post-processing via Levenberg-Marquardt fitting:

$$g^{(2)}(\tau) = 1 - a_1 e^{-\frac{\tau}{t_1}} + a_2 e^{-\frac{\tau}{t_2}}, \tag{3}$$

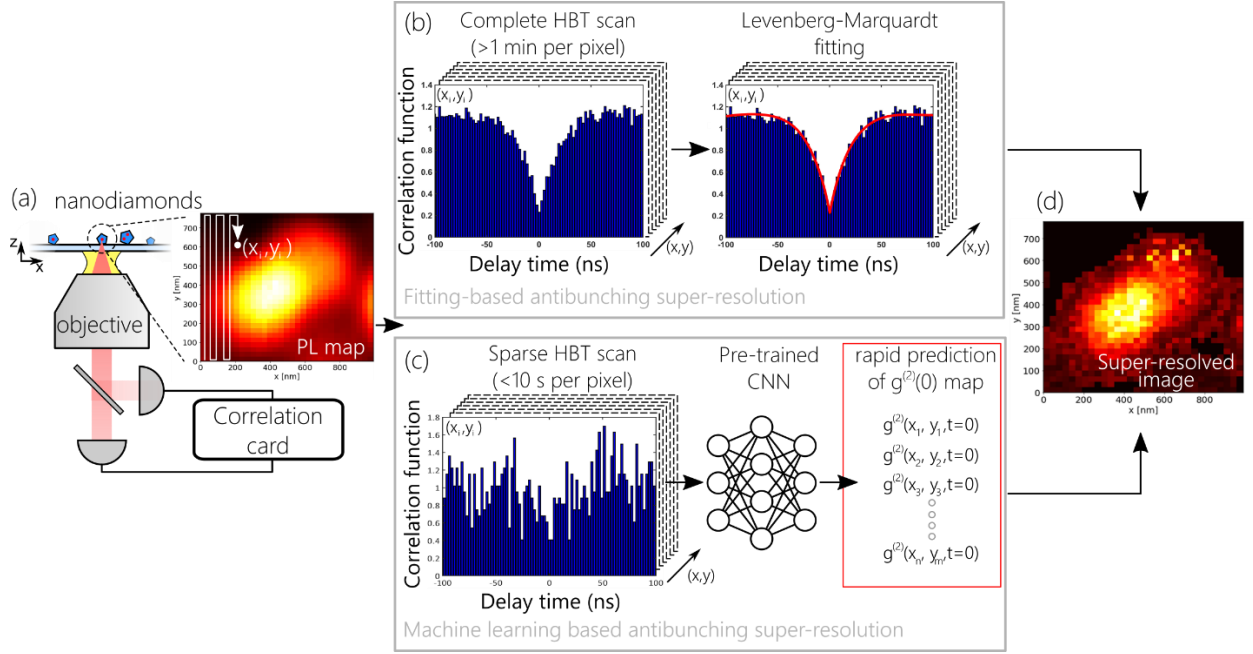

Figure 1. General framework of the machine learning (ML) assisted antibunching SRM. Antibunching-based SRM image acquisition starts with an area of n by m pixels (a) and measures complete antibunching histograms via Hanbury-Brown-Twiss (HBT) interferometry at each pixel (b). The Levenberg-Marquardt (L-M) fit is done on each pixel's HBT histogram to retrieve $g^{(2)}(x, y, 0)$ distribution. Finally, the super-resolved image is constructed using Eq. 2 (d). Alternatively, ML-assisted approach relies on pre-trained CNN regression model, which allows to accurately predict $g^{(2)}(x, y, 0)$ maps utilizing sparse HBT measurement data (c). The developed approach ensures at least 12 times speed-up compared with the conventional L-M fitting based antibunching SRM.

Here, $a_j$, $t_j$, $j = 1,2$ are the fitting parameters related to the internal dynamics of the emitters. Figure 1b shows the main steps of the fitting-based approach for the realization of the antibunching SRM technique. The area of interest is divided into $n \times m$ pixels, and autocorrelation histograms are acquired at each pixel. The autocorrelation measurement is performed for several minutes. The L-M fitting is done over all of the HBT histograms and the corresponding $g^{(2)}(x, y, 0)$ map is retrieved. Finally, the resolved image is calculated via Eq. (2) (Figure 1d).

In our demonstration, we use single nitrogen-vacancy (NV) centers in nanodiamonds dispersed on a coverslip glass substrate as single photon emitters. These emitters typically yield between $10^4$ and $10^5$ counts per second on each of the single-photon detectors in the HBT setup (when in focus) and exhibit fluorescence lifetimes between 10 and 100 ns. During the scan, when the emitters are partially out of focus, the fluorescence counts drop significantly. Consequently, in order to assess $\boldsymbol{g^{(2)}(0)}$ via Levenberg-Marquardt (L-M) fitting with an uncertainty varying between ±0.01 to ±0.05, autocorrelation histogram acquisition times of 1 min are required per pixel. In the pulsed excitation regime, the fitting is not required to retrieve $g^{(2)}(0)$ as long as the pump repetition period is much longer than the emitter's fluorescence lifetime. However, this requirement becomes somewhat impractical when the emitter lifetime is long as in the case of NV centers. The developed ML approach addresses the aforementioned problem by rapidly estimating the $\boldsymbol{g^{(2)}(x,y,0)}$ values based on sparse HBT measurement. The main framework of the developed approach is shown in Figure 1c. A CNN regression network is trained on a set of "sparse" autocorrelation data with short acquisition times. Once trained, the CNN network estimates the $g^{(2)}(0)$ values, requiring an acquisition time of less than 10 s.

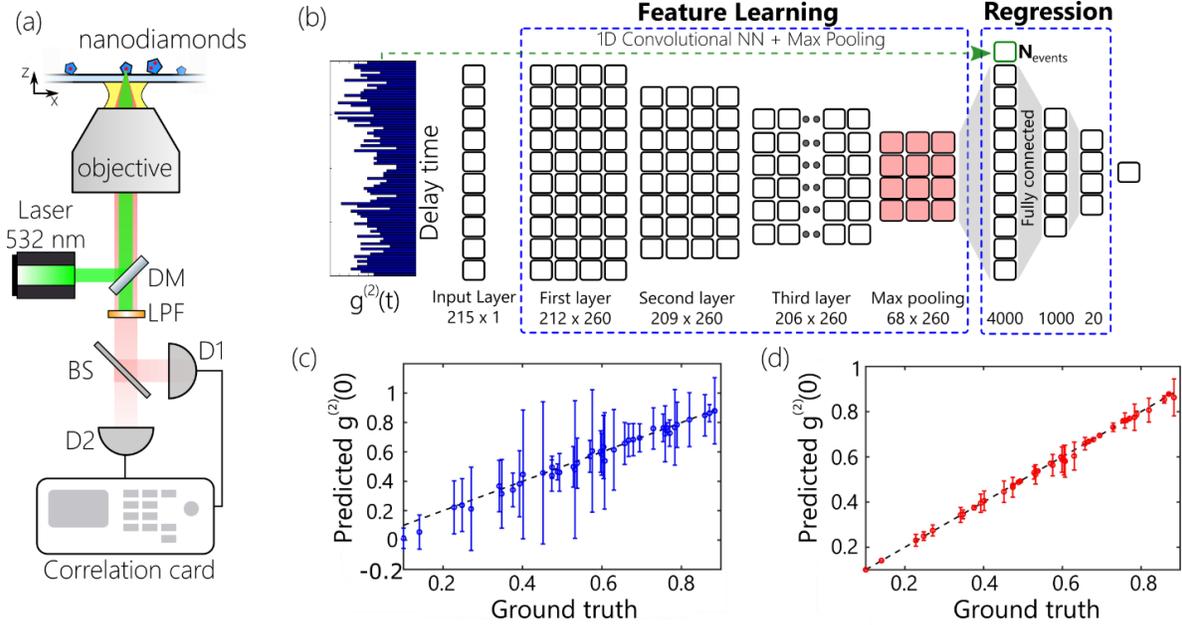

Figure 2. Machine learning assisted measurement of $g^{(2)}(0)$. (a) Schematics of the HBT interferometer. Labels: DM – dichroic mirror; LPF – long-pass filter; BS – beam splitter; D1/D2 – detectors. (b) Schematics of the CNN regression network. The input layer takes in sparse HBT histograms. The total number of events, $N_{events}$, of the histogram is concatenated to the output of the feature learning part and used as a regularization term. (c)-(d) Regression plot (predicted vs expected $\boldsymbol{g^{(2)}(0)}$ values) for L-M fitting (c) and CNN regression of $g^{(2)}(0)$ (d) from 5s datasets. Dots show the average predicted $g^{(2)}(0)$ value, while error bars shows the standard deviation of the predicted value over all the 5s datasets acquired for a given emitter.

## 3. Machine learning assisted autocorrelation function measurement

The main building block of our ML assisted antibunching SRM technique is the CNN based regression model, used for retrieving $\boldsymbol{g^{(2)}(0)}$ values. In this section, we highlight the structure of the CNN, its training and testing, as well as compare its performance against convention L-M fitting. The training dataset for sparse second-order autocorrelation histograms consists of measurements performed on a set of 40 randomly dispersed nanodiamonds with NV centers on a coverslip glass substrate. Figure 2a shows the schematics of the HBT setup used for these measurements. Two avalanche detectors (D1, D2) with 30 ps jitter are connected to a pulse correlator with a 4 ps internal jitter. The co-detection events are recorded over a range of 500 ns and collected into 215 equally sized time bins. For each of the 40 emitters, hundreds of sparse autocorrelation histograms with 1s acquisition time are collected, until the total number of co-detection events in their sum allows a precise ground truth ($\boldsymbol{g^{(2)}(0)}$) estimation via L-M fitting with fitting uncertainty varying between ±0.01 to ±0.05. The estimated ground truth value is then assigned as a label to the entire set of 1s histograms. We then formed all the possible combinations of 1 to 10 of these 1s histograms to obtain training data that emulated histograms with acquisition times from 1s to 10s. Such a data augmentation process assumed that the emission is a process with no memory over times exceeding 1s and allowed us to significantly extend the training dataset. More information on the training dataset collection process and data augmentation can be found in Supplementary materials.

Figure 2b shows the structure of the CNN used for $\boldsymbol{g^{(2)}(0)}$ regression. The CNN consists of one input layer, three hidden convolutional layers, one max-pooling layer followed by dropout, three fully connected layers, and one output node containing the regression result. The input layer had 215 nodes corresponding to the number of bins in the input histogram. The feature learning part of the CNN is optimized to capture the salient features of the autocorrelation datasets, while the regression part is trained to predict $\boldsymbol{g^{(2)}(0)}$ values based on these extracted features. All of the hidden layers comprised 260 filters. The third hidden layer's output is connected with the max-pooling layer, followed by the dropout layer. The kernel size of the filters (4) is chosen to be the same for each layer. Importantly, the CNN takes the total number of two-photon detection events $N_{events}$ in the histogram as an additional input. $N_{events}$ is concatenated to the output of the feature learning part and used as a regularization term during the training process. The 5s-10s histograms acquired on pixels where the contribution of the quantum emission to the total counts is negligible, feature $N_{events}<4$, while the histograms on areas close to the quantum emitter locations feature $N_{events}=65$ on average. To populate the "dark" pixels, the CNN regression network is implicitly biased to produce $\boldsymbol{g^{(2)}(0) = 1}$, on the datasets with $N_{events}<4$ counts. Supervised training of the CNN regression model was performed using the augmented dataset of 5s-10s sparse HBT histograms and the corresponding ground truth labels. The training process is realized by performing adamax gradient descent optimization using the Keras library [25] for 100 epochs with mean absolute percentage error loss function. 80% of the dataset is used for training, while the remaining 20% are used for validation and testing.

The performance of the trained CNN regression model is assessed via calculating the mean absolute percentage error (MAPE) and the coefficient of determination ($r^2$) on the 5s histogram datasets. Figure 2c shows the regression plot of the L-F fitting performed on 5s HBT histograms.

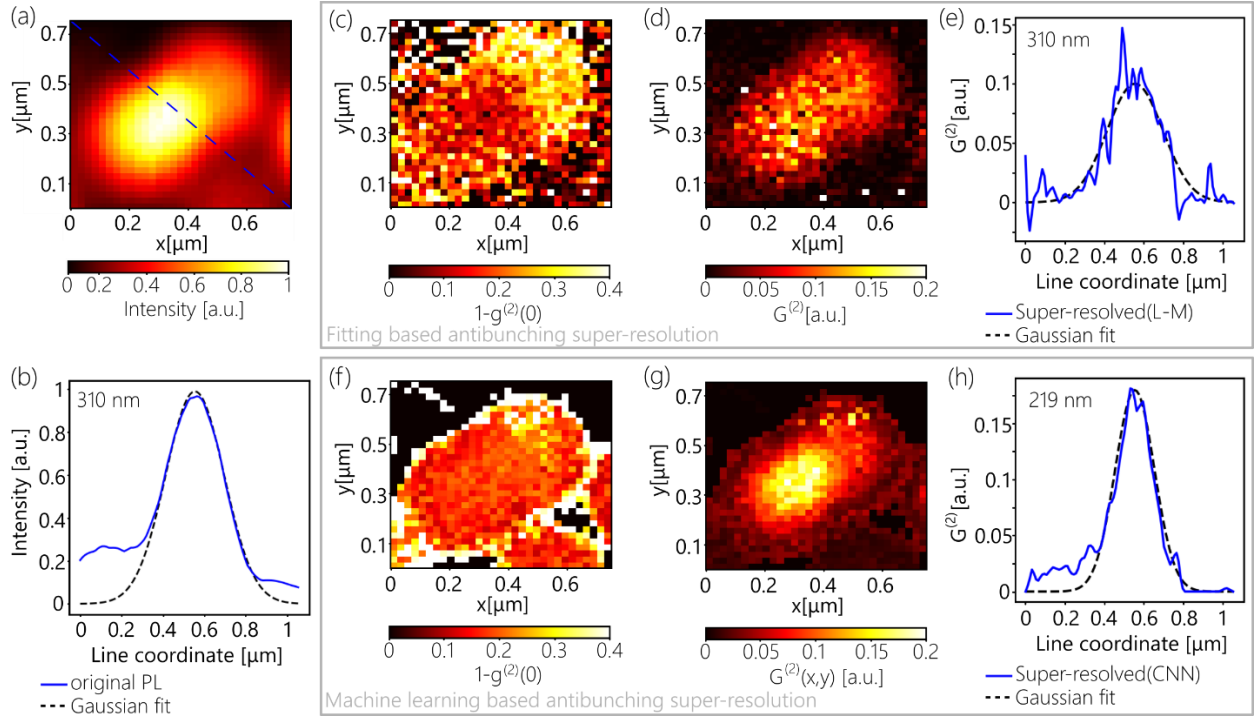

Figure 3. **Machine learning assisted antibunching SRM of a single NV center** (a). Photoluminescence (PL) distribution within the area of 32 by 32 pixels containing one NV center. (b) Cross section of the PL image (blue) along the dashed line in (a) and Gaussian fitting (dashed, black) with 310 nm FWHM. (c)-(e) Results of L-M based antibunching SRM based on 7s HBT measurement: distribution of retrieved $1 - g^{(2)}(x,y,0)$ (c); reconstructed image (d) and cross-section of G(2)(x,y) distribution of the L-M fitting based image (blue) and Gaussian fitting (dashed, black) with 310 nm FWHM (e). (f)-(h) Results of ML assisted antibunching SRM: $1 - g^{(2)}(x,y,0)$ map (f); reconstructed image (d); and corresponding cross section of intensity distribution of reconstructed image (blue) and the Gaussian fitting (dashed, black) with 219 nm FWHM.

Markers show the average value of the prediction, while error bars show the standard deviation over the set of 5s histograms belonging to the same emitter. Due to the sparsity of the HBT measurement, the L-M fitting expectedly cannot ensure precise fitting of the data, which results in MAPE=32%, $r^2$=70% and root mean square error (RMSE) of 0.215.

In contrast, the CNN regression model ensures very precise predictions of the $g^{(2)}(0)$ values based on 5s HBT histograms (Figure 2d). Due to the ability of the CNN network to learn hidden correlations between signature features of the sparse datasets and the ground truth labels, the CNN regression model shows excellent performance on the sparse dataset and ensures low MAPE (5%), a high coefficient of determination of 93% and RMSE of 0.0018. The CNN performance is also robust against the reduction of the acquisition time. We analyze the performance of both approaches on 5s, 6s and 7s HBT datasets. The performance of the direct fitting ensures 30% and 27% MAPE when applied to 6s and 7s HBT measurements, respectively.

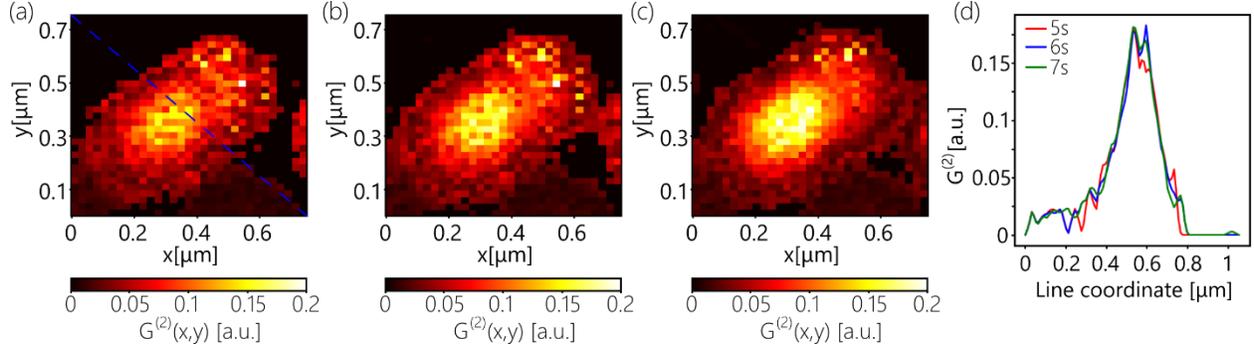

**Figure 4. Robustness of the machine learning assisted SRM against the reduction of acquisition time.** (a)-(c) Resolved images obtained via applying ML assisted antibunching SRM on 5s (a), 6s (b) and 7s (c) sparse HBT scans. Blue, dashed line shows the cross-section line. (d) Comparison of intensity cross-sections for three cases.

The CNN regression model ensures performance that is much more robust than L-M fitting. It ensures 3.92% MAPE on 6s HBT datasets and reaches up to 3.58% MAPE when applied to 7s datasets.

## 4. Experimental realization of machine learning assisted antibunching super-resolution microscopy

The benchmarking of the ML-assisted regression of autocorrelation data enables the experimental demonstration of the ML-assisted antibunching SRM. The experiment is realized on a sample of randomly dispersed nanodiamonds with NVs on a glass substrate. In this demonstration, the objective is scanned using a piezo-stage with sub-10 nm resolution over the 775x775 nm² region of interest, which is divided into 1024 (32x32) pixels and contains one nanodiamond with a single NV center. Autocorrelation measurements are performed on each pixel in 1s time increments with a 7s total acquisition time per pixel. Along with the autocorrelation data, the corresponding photoluminescence (PL) map is retrieved (Fig. 3a).

The cross-section of the diffraction-limited image, taken along the blue dashed line (Figure 3a), is shown in Figure 3b. Gaussian fitting of the intensity distribution yields a full width half maximum (FWHM= $2\sqrt{2\ln 2}\,\sigma$) of 310 nm. By L-M fitting the 5s sparse histograms of each pixel, the $g^{(2)}(x, y, 0)$ map is retrieved. Due to the sparsity of the HBT histograms, the L-M fitting expectedly leads to a noisy reconstruction of the $g^{(2)}(x, y, 0)$ distribution (Figure 3c). Figure 3d shows the corresponding reconstructed image of $G^{(2)}(x, y)$ (Eq. 3). The cross-section of the obtained image and corresponding fitting with the same $\sigma$ value as of the original PL image are shown in Figure 3e. Here we can see that the $g^{(2)}(x, y, 0)$ obtained via L-M fitting leads to a noisy, blurred image without any gain in spatial resolution, which is a direct consequence of the inaccurate retrieval of the $g^{(2)}(x, y, 0)$. In contrast, the CNN-based antibunching SRM ensures the expected $\sqrt{2}$ gain in resolution on sparse 7s HBT scan. Figure 3 (f-g) show $g^{(2)}(x, y, 0)$ distribution retrieved via using the pre-trained CNN (f) and corresponding super-resolved image (g). Here, we can see that ML-based framework ensures precise reconstruction of the $g^{(2)}(x, y, 0)$ map, and as a result achieves a $\sqrt{2}$ gain in the spatial resolution of the reconstructed image. Gaussian fitting of the cross-section distribution of the resolved image shows that ML assisted approach ensures a FWHM of 219 nm, which corresponds to $\boldsymbol{\sigma_{CNN} = \sigma/\sqrt{2}}$ (Figure 3h).

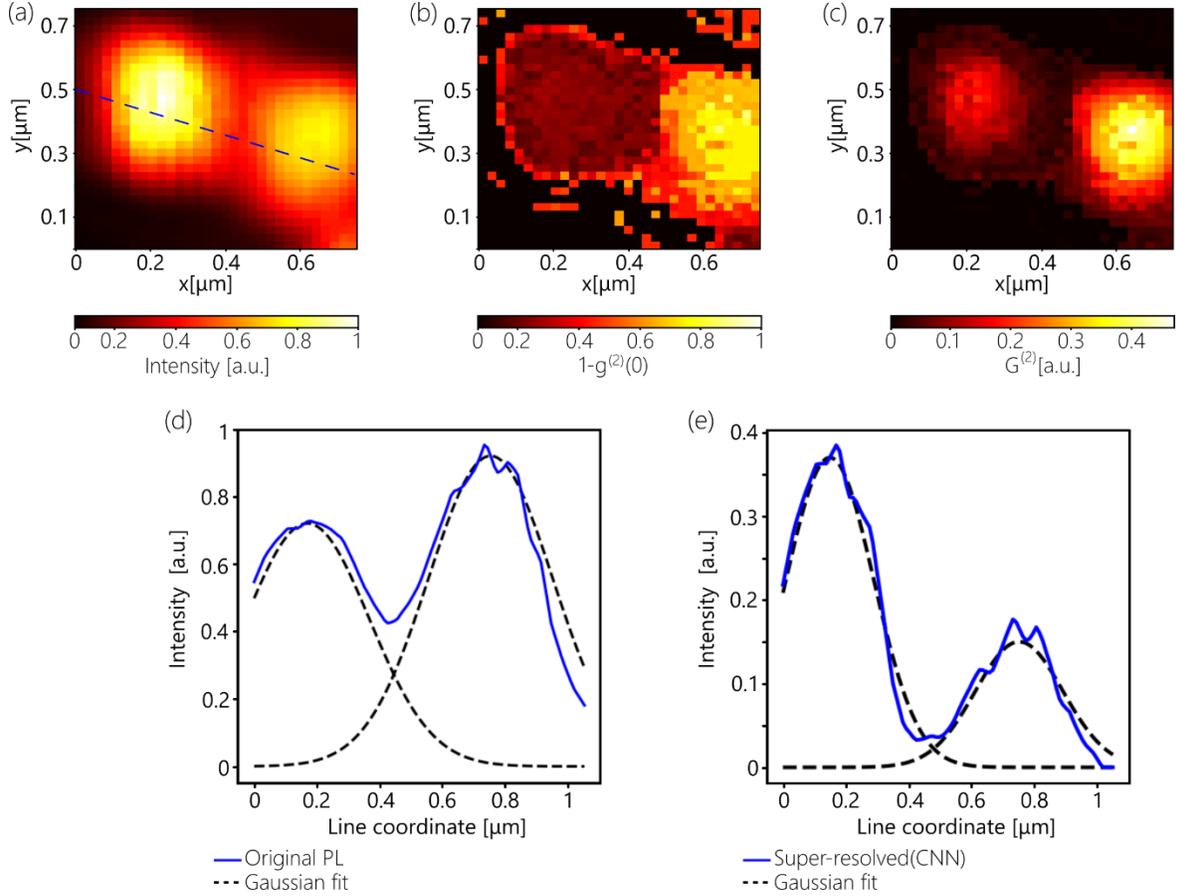

Figure 5. Machine learning assisted antibunching SRM resolves two closely spaced emitters. (a)-(c) Results of ML assisted antibunching SRM done on 7s HBT measurement: PL image (a); distribution of retrieved $\mathbf{1 - g^{(2)}(x, y, 0)}$ (b) and the reconstructed image (c). (d) cross-section of intensity distribution of the PL image(blue) and Gaussian fitting (dashed, black). Blue dashed line in (a) shows the cross-section line. (e) Cross-section of intensity distribution of the resolved image (blue) and Gaussian fitting (dashed, black).

Up till now, we have considered an acquisition time of 7s per pixel. However, the robustness of the regression model indicates that the developed approach can be efficiently applied to more sparse datasets. Figure 4(a)-(c) shows the reconstructed images based on 5s, 6s and 7s HBT scans, respectively, and Figure 4(d) compares their cross-sections, which appear stable against the reduction of the acquisition time. It is worth noting that the fitting-based approach requires at least 1 min of HBT measurement per pixel for precise retrieval of the $\boldsymbol{g^{(2)}(0)}$ values, as it has been observed during dataset collection process (Section 2). This time requirement significantly depends on the properties of the single-photon emitters, e.g. quantum purity, lifetime, and emission rate, and can be significantly longer in the case of low emission rates of the emitter.

Here, the developed ML-assisted anti-bunching approach ensures up to 12 times speed-up compared with the fitting-based approach.

The developed ML-assisted SRM is also capable of resolving closely spaced quantum emitters (Fig. 5). Figure 5(a)-(c) shows the PL distribution, CNN-based retrieved $\boldsymbol{g^{(2)}(x,y,0)}$ map and the resolved image of the two NVs separated by ~600 nm distance. By comparing the original PL distribution and the resolved image, the expected $\sqrt{\boldsymbol{2}}$ improvement in the spatial resolution is observed. By performing the Gaussian fitting of the cross-section (taken along the dashed line in Figure 5a), one can retrieve the FWHM values of each of the lobs, which are equal to ~465 nm (Figure 5d). By performing the same fitting on the resolved image, $\sqrt{\boldsymbol{2}}$ narrowing of the emission features (FWHM=330 nm) by the CNN based approach is confirmed.

## Conclusion

The proposed ML assisted regression technique allows for a significant speed up of quantum SRM imaging. Specifically, the performance of the CNN-assisted SRM is demonstrated on nanodiamonds which contain single NV centers as quantum emitters. In the microscopy of quantum light sources, the developed ML-assisted super-resolution framework ensures a speed-up of 12 times compared to the conventional L-M fitting-based approach for retrieving the second-order autocorrelation value at zero delays, $\boldsymbol{g^{(2)}(0)}$. The proposed approach can be extended to rapid measurements of higher-order autocorrelation functions, which opens up the way to practical realization of scalable quantum super-resolution imaging systems. The developed method is compatible with the CW excitation regime, which reduces emitter photobleaching due to multi-photon absorption and does not impose restrictions on the fluorescence lifetime. Therefore, it can be extended and applied to a wide variety of quantum emitters used in biological labeling, quantum on-chip photonics and beyond.

## Acknowledgment

This work is supported by the U.S. Department of Energy (DOE), Office of Science through the Quantum Science Center (QSC), a National Quantum Information Science Research Center (developing ML algorithms), DARPA/DSO Extreme Optics and Imaging (EXTREME) Program (HR00111720032) (A.V.K.) and National Science Foundation award 2029553-ECCS (sample fabrication).

# Supplementary materials for "Machine learning assisted quantum super-resolution microscopy"

Zhaxylyk A. Kudyshev[1,2], Demid Sychev[1,2], Zachariah Martin[1,2], Simeon I. Bogdanov[3,4], Xiaohui Xu[1,2], Alexander V. Kildishev[1], Alexandra Boltasseva[1,2], Vladimir M. Shalaev[1,2]

[1] School of Electrical and Computer Engineering, Birck Nanotechnology Center and Purdue Quantum Science and Engineering Institute, Purdue University, West Lafayette, IN, 47907, USA
[2] The Quantum Science Center (QSC), a National Quantum Information Science Research Center of the U.S. Department of Energy (DOE), Oak Ridge, TN 37931, USA
[3] Department of Electrical and Computer Engineering, University of Illinois at Urbana-Champaign, IL 61801, USA
[4] Nick Holonyak, Jr. Micro and Nanotechnology Laboratory, University of Illinois at Urbana-Champaign, IL 61801, USA

**Experimental setup.** The sample with nanodiamonds containing NV centers was prepared by cleaning a coverslip glass substrate with solvents, treating it with ultraviolet radiation for an hour, and drying a 5 µL droplet of a sonicated nanodiamond solution (20 nm average size, Adamas Nano) on the coverslip surface. Optical characterization was performed using a custom-made scanning confocal microscope with a 100 µm pinhole based on a commercial inverted microscope body (Nikon Ti–U). To locate the emitters, objective scanning was performed using a P-561 piezo stage driven by an E-712 controller (Physik Instrumente). The optical pumping in the CW experiments was administered by a continuous wave 532 nm laser (RGB Photonics). Power on the order of 1 mW (measured before entering the optical objective) was used to pump the NV centers. The excitation beam was reflected off a 550 nm long-pass dichroic mirror (DMLP550L, Thorlabs), and a 550 nm long-pass filter (FEL0550, Thorlabs) was used to filter out the remaining pump power. Two avalanche detectors with a 30 ps time resolution and 35% quantum efficiency at 650 nm (PDM, Micro-Photon Devices) were used for single-photon autocorrelation measurements. Time-correlated photon counting was performed by a "start-stop" acquisition card with a 4 ps internal jitter (SPC-150, Becker & Hickl). The total histogram span was set to 500 ns and the co-detection events were collected into 215 time bins.

**Training dataset.** In order to train the regression network, autocorrelation measurements were performed on a set of 40 emitters. For each emitter, autocorrelation datasets were acquired in series of 1-second-long intervals. These "sparse" datasets acquired for each emitter were compounded into a "full" dataset, from which the $g^{(2)}(0)$ value was attained using the L-M fitting algorithm. Autocorrelation measurements on each emitter were performed by repeating acquisitions for one second, until accumulating about 300 co-detection events per bin in total. To extract an estimate of the autocorrelation at zero delays, the complete autocorrelation histograms were fitted according to a three-level emitter model using the Levenberg-Marquardt (L-M) fit by the following function:

$$g^{(2)}(\tau) = 1 - a_1 e^{-\frac{\tau}{t_1}} + a_2 e^{-\frac{\tau}{t_2}}, \tag{S1}$$

The training dataset at this point consisted of 9416 sparse HBT histograms. The emitters in the dataset covered a broad range of $\boldsymbol{g^{(2)}(0)}$ values from 0.1 to 0.884, while the total number of counts of the 1s HBT histograms was in the range of 1.2 to 61.

The 1s HBT histograms were used for data augmentation. Specifically, we formed all the possible combinations of 1 to 10 of these 1s histograms to obtained training data that emulated histograms with acquisition times from 1s to 10s. This was done via bin-wise summation of the histograms. Such data augmentation process assumed that the emission is a process with no memory over times exceeding 1s and allowed us to significantly extend the training dataset.

**Convolutional neural network structure and training process.** A 1D CNN was used to implement the regression model. Table S1 shows a detailed summary of the regression network structure. Note that an additional input parameter (total number of counts) is concatenated with the output of the feature learning part of the network. This concatenation is performed to introduce additional information into the network to ensure better training outcomes on the HBT datasets with different acquisition times but the same $\boldsymbol{g^{(2)}(0)}$ values.

The CNN model consisted of one input layer, three hidden convolutional layers ("relu" activation function), one max-pooling and dropout layer. The output of the dropout layer is used as an input into the fully connected layer, which consisted of four dense layers and three dropouts. Mean absolute percentage error was used as an error loss function, while the adamax gradient descent optimization was used for training the network. The regression network was trained on 80% of the extended 5s-10s HBT dataset for 100 epochs, while the remaining 20% of the data was used for validation/testing. Figure S1a shows the evolution of the training/validation losses during the training phase.

| Layer type | Output shape | Number of parameters | Connected to |
|---|---|---|---|
| Input_1 | [215, 1] | | |
| Conv1D_1 | [212, 260] | 1300 | Input_1 |
| Conv1D_2 | [209, 260] | 270660 | Conv1D_1 |
| Conv1D_3 | [206, 260] | 270660 | Conv1D_2 |
| MaxPooling1D | [68, 260] | 0 | Conv1D_3 |
| Dropout | [68, 260] | 0 | MaxPooling1D |
| Input_2 | [1] | 0 | |
| Flatten | 17680 | 0 | Dropout |
| Concatenate | 17681 | 0 | Flatten, Input_2 |
| Dense_1 | 8041 | 142180962 | Concatenate |
| Dropout_1 | 8041 | 0 | Dense_1 |
| Dense_2 | 4000 | 32168000 | Dropout_1 |
| Dropout_2 | 4000 | 0 | Dense_2 |
| Dense_3 | 1000 | 4001000 | Dropout_2 |
| Dropout_3 | 1000 | 0 | Dense_3 |
| Dense_4 | 20 | 20020 | Dropout_3 |
| Dense_5 | 1 | 21 | Dense_4 |

Table S1. Structure of the CNN regression network.

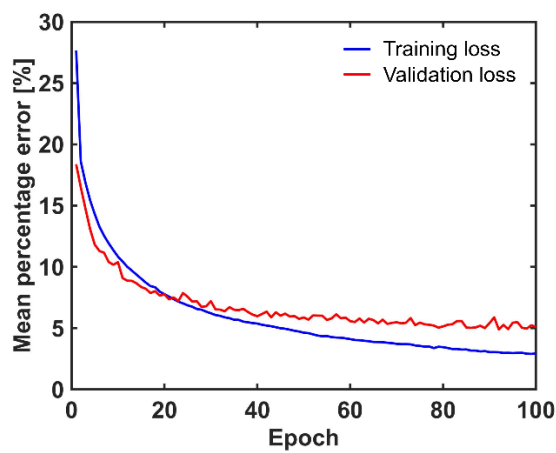

Figure S1. Evolution of the training and validation loss of the CNN regression network during the training.